\documentstyle{elsart}
\input psfig

\begin{document}
\begin{frontmatter}
\hspace{2.0in} {\rm \hfill OHSTPY-HEP-E-97-009}

\title{Incorporation of the statistical uncertainty in 
the background estimate into the upper limit on the signal}
\author{K.K.~Gan, J.~Lee, R.~Kass}
\address{Department of Physics,
        The Ohio State University,
        Columbus OH 43210,
        U.S.A.}

\begin{abstract}
We present a procedure for calculating an upper limit on the
number of signal events which incorporates the Poisson
uncertainty in the background, estimated from  control
regions of one or two dimensions. 
For small number of signal events, the upper limit obtained
is more stringent than that extracted without including the
Poisson uncertainty.
This trend continues until the number of background events
is comparable with the signal.
When the number of background events is comparable or larger
than the signal, the upper limit obtained is less stringent
than that extracted without including the Poisson uncertainty.
It is therefore important to incorporate the Poisson uncertainty
into the upper limit; otherwise the upper limit obtained
could be too stringent.
\end{abstract}
\end{frontmatter}

\section{Introduction}

In the search for a rare or forbidden process, an experiment usually
observes zero or few candidate events and sets an upper limit on the
number of signal events.
The background is often estimated from 
control regions of one or two dimensions.
In the case of a small number of candidates events, it is
important to incorporate the uncertainty in the background
estimate due to Poisson statistics into the upper limit. 

For $N_{ob}$ observed events with an expected background of
$N_{bg}$ events, the upper limit on the number of signal events
$\lambda_0$ at a confidence level $\delta$ is given by:
\begin{eqnarray}
\label{eq:exact}
\frac 
{ e^{-(\lambda_0 + N_{bg}) } 
\sum_{n=0}^{N_{ob}} \frac{(\lambda_0 + N_{bg})^n}{n!} }
{ e^{-N_{bg}} \sum_{n=0}^{N_{ob}} \frac{(N_{bg})^n}{n!} }
= 1-\delta,
\end{eqnarray}
assuming that there is no uncertainty in the background
estimate~\cite{Zech,PDG}.
In this paper, we present a procedure for incorporating the statistical
uncertainty in the background estimate using Poisson statistics.

\section{Incorporation of the statistical uncertainty}

\begin{figure}[t]
\centering
\centerline{\hbox{\psfig{figure=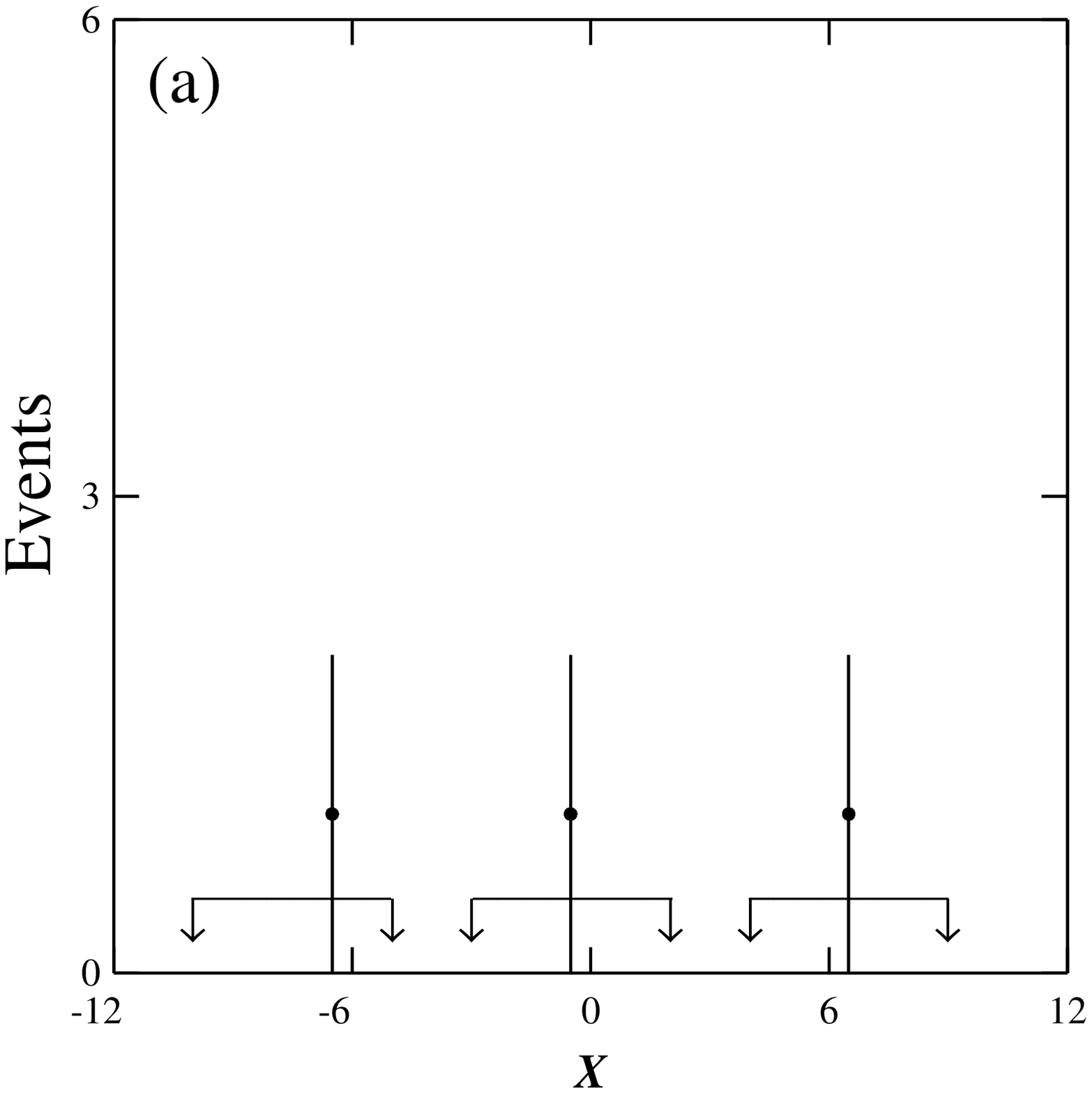,width=2.5in,height=2.5in}
\hspace{0.2in}
\psfig{figure=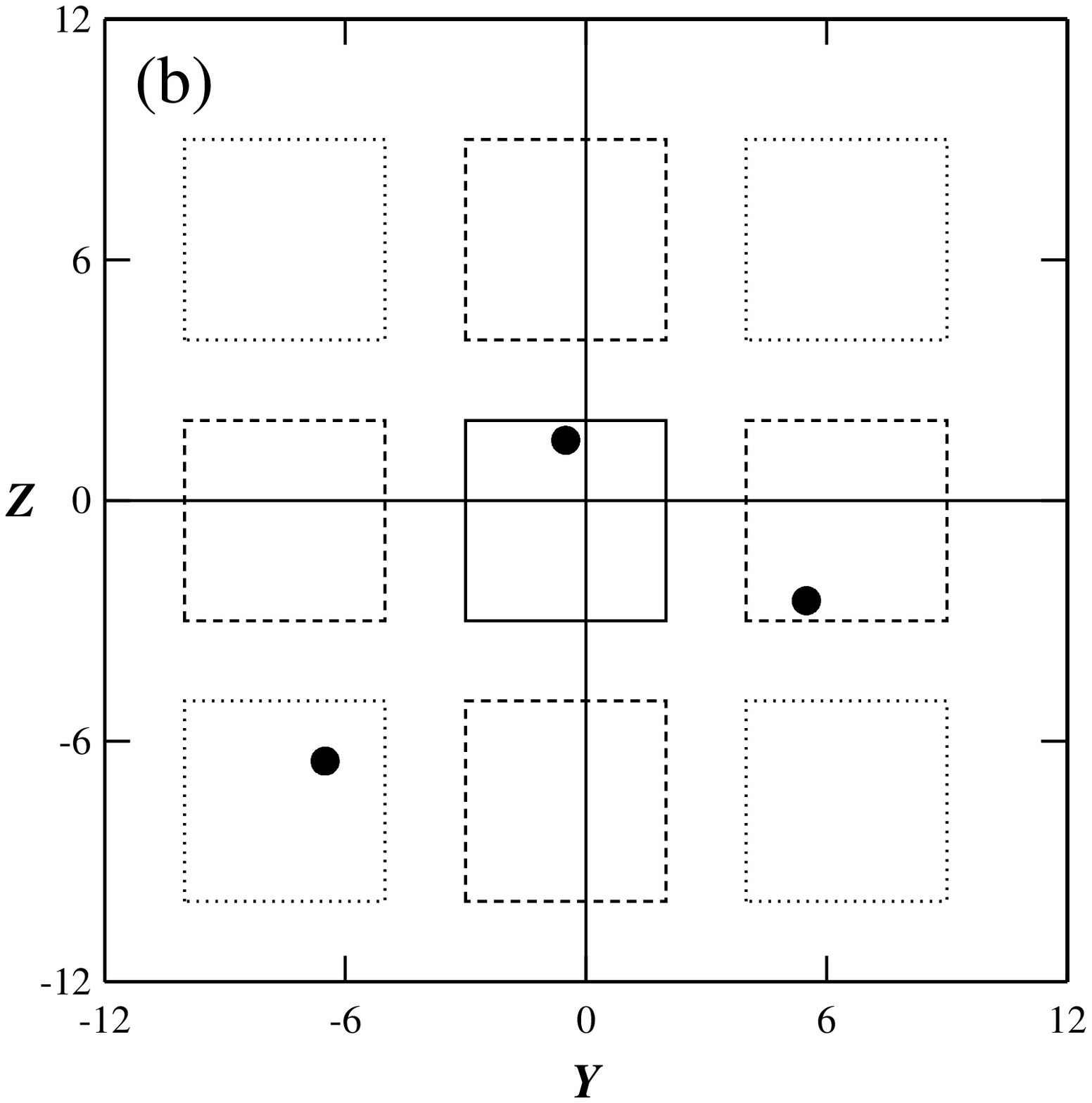,width=2.5in,height=2.5in}}}
\caption{(a) Definition of signal and two sideband regions 
(arrowed brackets) in the $X$ distribution for the
one-dimensional case.
The width of the sideband regions is the same as the signal region.
(b) Definition of signal, sideband (dashed), and corner-band
(dotted) regions in the $Y$ vs. $Z$ distribution for the
two-dimensional case.
The area of a sideband or corner-band region is the same
as the signal region.}
\label{fig:example1}
\end{figure}

We consider the case in which the background is estimated from 
control regions of one or two dimensions which have limited statistics.
Figure~\ref{fig:example1}(a) shows an example of a signal and two
sideband regions in the distribution of a physical variable
for the one-dimensional case.
The corresponding example for the two-dimensional case
for the signal, four sideband and four corner-band
regions is shown in Fig.~\ref{fig:example1}(b).
Assuming that the background is linear in the vicinity of the
signal region, the background in the signal region can be estimated
from the number 
of events in the sideband (and corner-band if appropriate) regions.
The number of signal events $N_0$ is given by:
\begin{eqnarray}
\label{eq:signal}
N_0 = N_{ob} - N_{bg} \equiv N_{ob} - \alpha N_{sb} + \beta N_{cb},
\end{eqnarray}
where $N_{sb}$ ($N_{cb}$) is the number of events
in the sideband (corner-band) regions.
For the one-dimensional case, $\alpha$ is the ratio of the width
of the signal region to the total width of the sideband regions
and $\beta$ is zero.
In the two-dimensional case, $\alpha$ is two times the ratio of
the area of the signal region to the total area of the sideband
regions and $\beta$ is the ratio of the area of the signal region
to the total area of the corner-band regions. 

The extraction of the upper limit on the signal must
account for the uncertainty in the background estimate due to limited
statistics in the control regions.  
For the one-dimensional case, this is implemented by expanding
Eq.~(\ref{eq:exact}) to sum over all possible background values
weighted by the Poisson probability for observing the $N_{sb}$
background events given an expected number of events
$\lambda_{sb}$ in the sideband regions.
The upper limit on the number of signal events $\lambda$ at a
confidence level $\delta$
is given by:
\begin{eqnarray}
\label{eq:1d}
\nonumber 
&&
\frac {  
\int_0^{\infty} d\lambda_{sb}
\frac {\lambda_{sb}^{N_{sb}}} {N_{sb}!} e^{-\lambda_{sb}}
e^{-(\lambda + \alpha\lambda_{sb})}
\sum_{n=0}^{N_{ob}} \frac{(\lambda + \alpha\lambda_{sb})^n}{n!}  
      }
      {  
\int_0^{\infty} d\lambda_{sb}
\frac {\lambda_{sb}^{N_{sb}}} {N_{sb}!} e^{-\lambda_{sb}}
     e^{-\alpha\lambda_{sb}}
\sum_{n=0}^{N_{ob}} \frac{(\alpha\lambda_{sb})^n}{n!}  
	     } \\
\nonumber 
&=&
\frac { 
\sum_{n=0}^{N_{ob}} \sum_{k=0}^{n}
\frac {\alpha^k \lambda^{n-k} e^{-\lambda} (N_{sb}+k)!}
{(1+\alpha)^{k}k!(n-k)!}       
      }
      {
\sum_{k=0}^{N_{ob}}
\frac{\alpha^k (N_{sb}+k)!}{(1+\alpha)^{k}k!}  
      } \\
&=& 1-\delta. 
\end{eqnarray}
The integral of $\lambda_{sb}$ is performed with 
$\int_0^{\infty} dy e^{-y} y^k = k!$.
Note that this formula can also be used to calculate the upper
limit on the number of signal events for the case in which 
the background is estimated
from a Monte Carlo simulation.
In this case, $N_{sb}$ is the number of background events that
satisfied the event selection criteria in a Monte Carlo sample
that is $\alpha^{-1}$ times larger than the data.

For control regions of two dimensions, the formula for the upper
limit is much more complicated.
However, it can be simplified depending on the value of $\beta$.
In the case of $\beta=1$ which corresponds a total corner-band area
that is equal to the area of the signal region, the upper limit is
given by:
\begin{eqnarray}
\label{eq:2d1}
\nonumber 
&&
\frac {  
\int_0^{\infty} d\lambda_{sb}
\int_0^{\alpha\lambda_{sb}} d\lambda_{cb}
\frac {\lambda_{sb}^{N_{sb}}} {N_{sb}!} e^{-\lambda_{sb}}
\frac {\lambda_{cb}^{N_{cb}}} {N_{cb}!} e^{-\lambda_{cb}}
e^{-(\lambda+\alpha\lambda_{sb}-\lambda_{cb})}
\sum_{n=0}^{N_{ob}} \frac {
(\lambda+\alpha\lambda_{sb}-\lambda_{cb})^n} {n!}
      }
      {  
\int_0^{\infty} d\lambda_{sb}
\int_0^{\alpha\lambda_{sb}} d\lambda_{cb}
\frac {\lambda_{sb}^{N_{sb}}} {N_{sb}!} e^{-\lambda_{sb}}
\frac {\lambda_{cb}^{N_{cb}}} {N_{cb}!} e^{-\lambda_{cb}}
e^{-(\alpha\lambda_{sb}-\lambda_{cb})}
\sum_{n=0}^{N_{ob}} \frac {(\alpha\lambda_{sb}-\lambda_{cb})^n} {n!}
      } \\ 
\nonumber 
&=&
\frac { 
\sum_{n=0}^{N_{ob}} \sum_{k=0}^n
\frac {\alpha^k \lambda^{n-k} e^{-\lambda}} {(1+\alpha)^k (n-k)!}
\sum_{j=0}^k 
\frac {(-1)^{k-j} (N_{sb}+N_{cb}+k+1)!} 
      {j!(k-j)! (N_{cb}+k-j+1)}
      }
      {
\sum_{k=0}^{N_{ob}} \frac{\alpha^k}{(1+\alpha)^k}
\sum_{j=0}^k 
\frac {(-1)^{k-j} (N_{sb}+N_{cb}+k+1)!} 
	{j!(k-j)! (N_{cb}+k-j+1)}
      } \\
&=& 1-\delta,
\end{eqnarray}
where $\lambda_{cb}$ is the expected number of events
in the corner-band regions. 
Note that the unphysical region ($\lambda_{cb} > \alpha \lambda_{sb}$)
is excluded from the integral.

For a larger corner-band area, $0<\beta<1$,
the upper limit is given by:
\begin{eqnarray}
\label{eq:2d2}
\nonumber 
&&
\frac {  
\int_0^{\infty} d\lambda_{sb}
\int_0^{\gamma\lambda_{sb}} d\lambda_{cb}
\frac {\lambda_{sb}^{N_{sb}}} {N_{sb}!} e^{-\lambda_{sb}}
\frac {\lambda_{cb}^{N_{cb}}} {N_{cb}!} e^{-\lambda_{cb}}
e^{-(\lambda+\alpha\lambda_{sb}-\beta\lambda_{cb})}
\sum_{n=0}^{N_{ob}} \frac {
(\lambda+\alpha\lambda_{sb}-\beta\lambda_{cb})^n} {n!}
      }
      {  
\int_0^{\infty} d\lambda_{sb}
\int_0^{\gamma\lambda_{sb}} d\lambda_{cb}
\frac {\lambda_{sb}^{N_{sb}}} {N_{sb}!} e^{-\lambda_{sb}}
\frac {\lambda_{cb}^{N_{cb}}} {N_{cb}!} e^{-\lambda_{cb}}
e^{-(\alpha\lambda_{sb}-\beta\lambda_{cb})}
\sum_{n=0}^{N_{ob}} 
\frac {(\alpha\lambda_{sb}-\beta\lambda_{cb})^n} {n!}
      } \\ 
\nonumber 
&=&
\frac {
\sum_{n=0}^{N_{ob}} \sum_{k=0}^n \alpha^k 
\frac {\lambda^{n-k} e^{-\lambda}} {(n-k)!} 
\sum_{j=0}^k 
\frac {(-1)^{k-j}(N_{cb}+k-j)!} {(\gamma-\alpha)^{k-j} j!(k-j)!}
\left[\frac {(N_{sb}+j)!} {(1+\alpha)^{N_{sb}+j+1}}-
\sum_{i=0}^{N_{cb}+k-j} 
\frac {(\gamma-\alpha)^{i}(N_{sb}+j+i)!} 
{(1+\gamma)^{N_{sb}+j+i+1} i!}\right]
      }
      {
\sum_{k=0}^{N_{ob}} \alpha^k
\sum_{j=0}^k 
\frac {(-1)^{k-j}(N_{cb}+k-j)!} {(\gamma-\alpha)^{k-j} j!(k-j)!}
\left[\frac {(N_{sb}+j)!} {(1+\alpha)^{N_{sb}+j+1}}-
\sum_{i=0}^{N_{cb}+k-j} 
\frac {(\gamma-\alpha)^{i}(N_{sb}+j+i)!} 
{(1+\gamma)^{N_{sb}+j+i+1} i!}\right]
      } \\
&=& 1-\delta,
\end{eqnarray}

where the unphysical region
($\lambda_{cb} > \frac{\alpha}{\beta} \lambda_{sb} 
\equiv \gamma \lambda_{sb}$)
is excluded from the integral.
The integral of $\lambda_{cb}$ is obtained with 
$\int_0^{x} dy e^{-y} y^k = k! (1-\sum_{i=0}^{k}\frac{x^i}{i!}e^{-x})$.

In the following sections, we consider two widely used
relative dimensions between the signal and control regions.

\subsection{Case I: $\alpha = \frac{1}{2}$ and $\beta = \frac{1}{4}$}

We first consider the one-dimensional case in which the total width
of the sideband regions is twice that of the signal region as
shown in Fig.~\ref{fig:example1}(a).
This corresponds to $\alpha = \frac{1}{2}$ and
Eq.~(\ref{eq:1d}) for the upper limit on the signal is simplified to:
\begin{eqnarray}
\label{eq:1d_1}
\nonumber 
&&
\frac {  
\int_0^{\infty} d\lambda_{sb}
\frac {\lambda_{sb}^{N_{sb}}} {N_{sb}!} e^{-\lambda_{sb}}
e^{-(\lambda + \frac{\lambda_{sb}}{2})}
\sum_{n=0}^{N_{ob}} \frac{(\lambda + \frac{\lambda_{sb}}{2})^n}{n!}  
      }
      {  
\int_0^{\infty} d\lambda_{sb}
\frac {\lambda_{sb}^{N_{sb}}} {N_{sb}!} e^{-\lambda_{sb}}
e^{-\frac{\lambda_{sb}}{2}}
\sum_{n=0}^{N_{ob}} \frac{(\frac{\lambda_{sb}}{2})^n}{n!}
      } \\
\nonumber 
&=&
\frac {
\sum_{n=0}^{N_{ob}}
\sum_{k=0}^{n} 
\frac {\lambda^{n-k} e^{-\lambda} (N_{sb}+k)! } {3^{k} k! (n-k)!}  
      }
      {
\sum_{k=0}^{N_{ob}} \frac{(N_{sb}+k)!}{3^{k} k!}  
      } \\
&=& 1-\delta. 
\end{eqnarray}

We now consider the two-dimensional case in which the total area
of both the sideband and corner-band regions is four times that
of the signal region as shown in Fig.~\ref{fig:example1}(b).
This corresponds to $\alpha = \frac{1}{2}$ and $\beta = \frac{1}{4}$.
From Eq.~(\ref{eq:2d2}), the upper limit on the signal is given by:
\begin{eqnarray}
\label{eq:2d2_1}
\nonumber 
&&
\frac {  
\int_0^{\infty} d\lambda_{sb}
\int_0^{2\lambda_{sb}} d\lambda_{cb}
\frac {\lambda_{sb}^{N_{sb}}} {N_{sb}!} e^{-\lambda_{sb}}
\frac {\lambda_{cb}^{N_{cb}}} {N_{cb}!} e^{-\lambda_{cb}}
e^{-(\lambda + \frac{\lambda_{sb}}{2} - \frac{\lambda_{cb}}{4})}
\sum_{n=0}^{N_{ob}} \frac{(\lambda + 
\frac{\lambda_{sb}}{2} - \frac{\lambda_{cb}}{4}
)^n}{n!}  
      }
      {  
\int_0^{\infty} d\lambda_{sb}
\int_0^{2\lambda_{sb}} d\lambda_{cb}
\frac {\lambda_{sb}^{N_{sb}}} {N_{sb}!} e^{-\lambda_{sb}}
\frac {\lambda_{cb}^{N_{cb}}} {N_{cb}!} e^{-\lambda_{cb}}
e^{-(\frac{\lambda_{sb}}{2} - \frac{\lambda_{cb}}{4})}
\sum_{n=0}^{N_{ob}} \frac{(\frac{\lambda_{sb}}{2} - 
\frac{\lambda_{cb}}{4})^n}{n!}
      } \\ 
\nonumber 
&=&
\frac {
\sum_{n=0}^{N_{ob}} 
\sum_{k=0}^{n} \frac {\lambda^{n-k} e^{-\lambda}} {3^{k}(n-k)!} 
\sum_{j=0}^{k} \frac {(-1)^{k-j} (N_{cb}+k-j)!} {j! (k-j)!}
[ (N_{sb}+j)! - \sum_{i=0}^{N_{cb}+k-j} \frac { (N_{sb}+j+i)! } 
{2^{N_{sb}+j+i+1}i!} ]
      }
      {
\sum_{k=0}^{N_{ob}} \frac{1}{3^{k}} 
\sum_{j=0}^{k} \frac {(-1)^{k-j} (N_{cb}+k-j)!} {j! (k-j)!}
[ (N_{sb}+j)! - \sum_{i=0}^{N_{cb}+k-j} \frac { (N_{sb}+j+i)! } 
{2^{N_{sb}+j+i+1}i!} ]
      } \\
&=& 1-\delta. 
\end{eqnarray}

\subsection{Case II: $\alpha = \beta = 1$}

\begin{figure}[t]
\centering
\centerline{\hbox{\psfig{figure=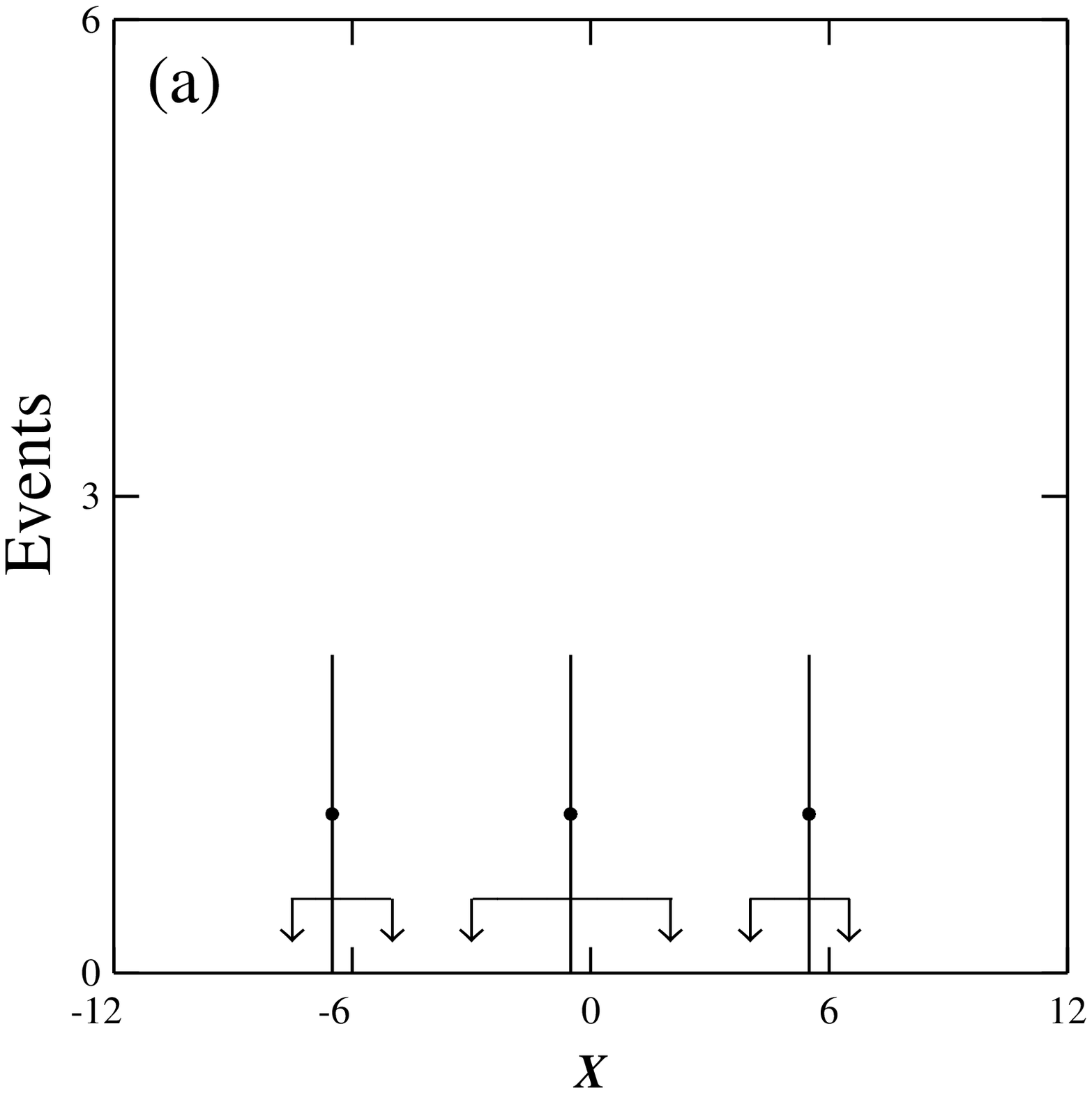,width=2.5in,height=2.5in}
\hspace{0.2in}
\psfig{figure=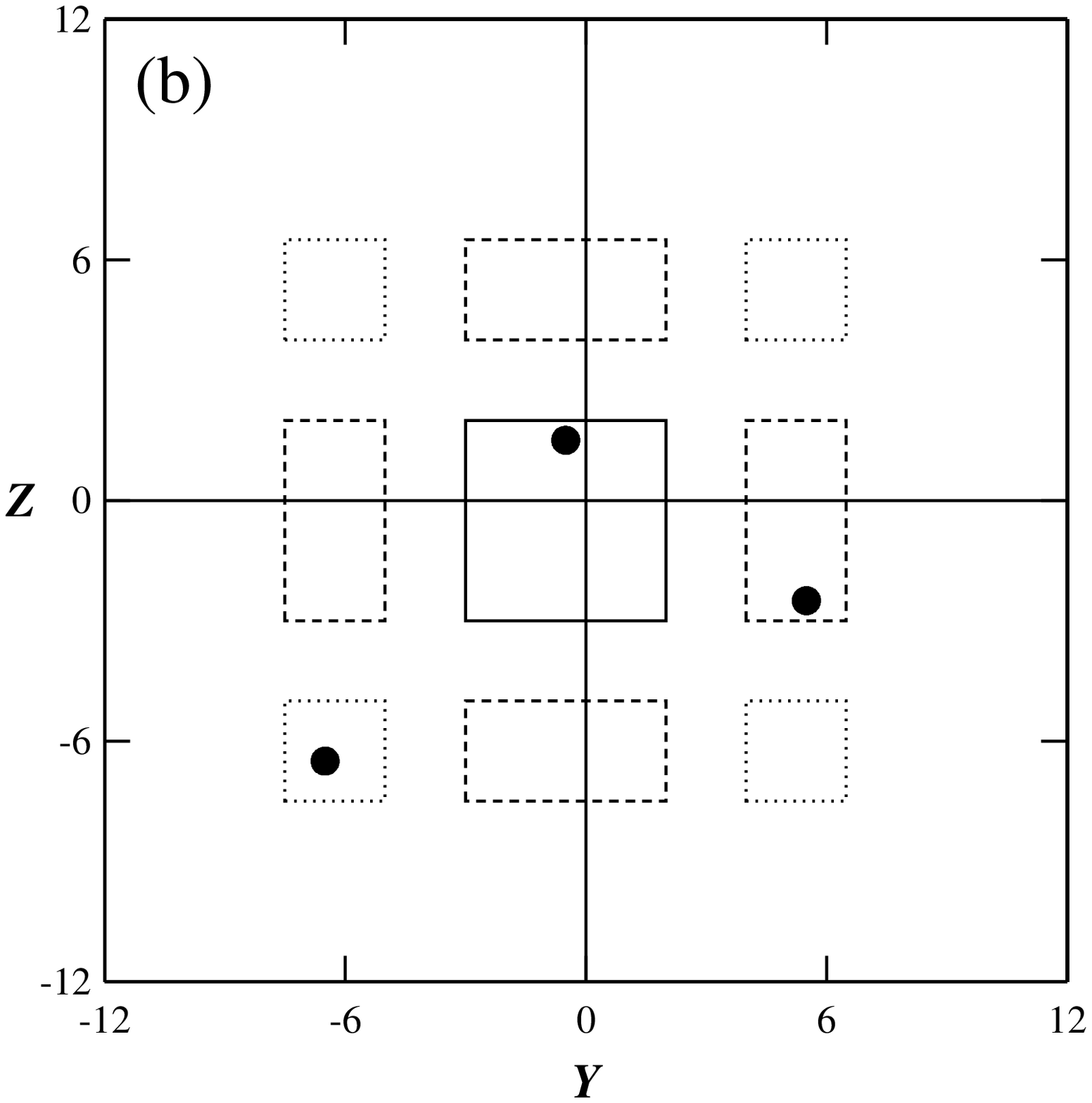,width=2.5in,height=2.5in}}}
\caption{(a) Definition of signal and two sideband regions 
(arrowed brackets) in the $X$ distribution for the
one-dimensional case.
The width of each sideband region is half that of the signal region.
(b) Definition of signal, sideband (dashed), and corner-band
(dotted) regions in the $Y$ vs. $Z$ distribution for the
two-dimensional case.
The area of each sideband (corner-band) region is
$\frac{1}{2}$ ($\frac{1}{4}$) that of the signal region.}
\label{fig:example2}
\end{figure}

We first consider the one-dimensional case in which the total width
of the sideband regions is the same as that of the signal region
(Fig.~\ref{fig:example2}(a)); an experimenter may choose
this kind of smaller background control regions so that the
background is more linear in the vicinity of the signal region.
Substituting for $\alpha = 1$ in Eq.~(\ref{eq:1d}), the upper limit
of the signal is given by:
\begin{eqnarray}
\label{eq:1d_2}
\nonumber 
&&
\frac {  
\int_0^{\infty} d\lambda_{sb}
\frac {\lambda_{sb}^{N_{sb}}} {N_{sb}!} e^{-\lambda_{sb}}
e^{-(\lambda + \lambda_{sb})}
\sum_{n=0}^{N_{ob}} \frac{(\lambda + \lambda_{sb})^n}{n!}  
      }
      {  
\int_0^{\infty} d\lambda_{sb}
\frac {\lambda_{sb}^{N_{sb}}} {N_{sb}!} e^{-\lambda_{sb}}
e^{-\lambda_{sb}}
\sum_{n=0}^{N_{ob}} \frac{(\lambda_{sb})^n}{n!}
      } \\
\nonumber 
&=&
\frac {
\sum_{n=0}^{N_{ob}}
\sum_{k=0}^{n} 
\frac {\lambda^{n-k} e^{-\lambda} (N_{sb}+k)! } {2^{k} k! (n-k)!}  
      }
      {
\sum_{k=0}^{N_{ob}} \frac{(N_{sb}+k)!}{2^{k} k!}  
      } \\
&=& 1-\delta. 
\end{eqnarray}

We now consider the two-dimensional case in which the total area of
the sideband regions is twice that of the signal region and the total
area of corner-band regions is the same as
that of the signal region (Fig.~\ref{fig:example2}(b)).
The upper limit is given by Eq.~(\ref{eq:2d1}) through substitution of
$\alpha = \beta = 1$:
\begin{eqnarray}
\label{eq:2d1_1}
\nonumber 
&&
\frac {  
\int_0^{\infty} d\lambda_{sb}
\int_0^{\lambda_{sb}} d\lambda_{cb}
\frac {\lambda_{sb}^{N_{sb}}} {N_{sb}!} e^{-\lambda_{sb}}
\frac {\lambda_{cb}^{N_{cb}}} {N_{cb}!} e^{-\lambda_{cb}}
e^{-(\lambda + \lambda_{sb} - \lambda_{cb})}
\sum_{n=0}^{N_{ob}}
 \frac{(\lambda + \lambda_{sb} - \lambda_{cb})^n}{n!}  
      }
      {  
\int_0^{\infty} d\lambda_{sb}
\int_0^{\lambda_{sb}} d\lambda_{cb}
\frac {\lambda_{sb}^{N_{sb}}} {N_{sb}!} e^{-\lambda_{sb}}
\frac {\lambda_{cb}^{N_{cb}}} {N_{cb}!} e^{-\lambda_{cb}}
e^{-(\lambda_{sb} - \lambda_{cb})}
\sum_{n=0}^{N_{ob}} \frac{(\lambda_{sb} - \lambda_{cb})^n}{n!}
      } \\ 
\nonumber 
&=&
\frac {
\sum_{n=0}^{N_{ob}} 
\sum_{k=0}^{n} \frac {\lambda^{n-k} e^{-\lambda}} {2^{k}(n-k)!} 
\sum_{j=0}^{k} \frac {(-1)^{k-j} (N_{sb}+N_{cb}+k+1)!}
 {j! (k-j)! (N_{cb}+k-j+1)}
      }
      {
\sum_{k=0}^{N_{ob}} \frac{1}{2^{k}} 
\sum_{j=0}^{k} \frac {(-1)^{k-j} (N_{sb}+N_{cb}+k+1)! }
 {j! (k-j)! (N_{cb}+k-j+1)}
      } \\
&=& 1-\delta.
\end{eqnarray}

\section{Results}

We have investigated the implication of including the
Poisson uncertainty in the estimated background in the
calculation of upper limits.
In Tables~\ref{tab:1d}-\ref{tab:2d2}, we give the 90\% and
95\% confidence level upper limits on the signal with and
without including the Poisson uncertainty for
several cases of small number of observed events
and various backgrounds.
The control regions have been chosen such that
$\alpha = \frac{1}{2}$ and $\beta = \frac{1}{4}$.
In the calculation of $\lambda_0$ with Eq.~(\ref{eq:exact}),
the background is set to zero if $N_{cb} > 2 N_{sb}$.
As expected, $\lambda$ is identical to $\lambda_0$ when the
number of events in the signal region is zero, regardless of the
number of estimated background events.
However, $\lambda$ is smaller than $\lambda_0$ when the number
of events in the signal region is non-zero and the estimated
background is zero.
This is not unexpected because Eqs.~(\ref{eq:1d_1}) and
(\ref{eq:2d2_1}) allow for the estimated zero background to
fluctuate up, in contrast to Eq.~(\ref{eq:exact}) in which the
background is estimated to be zero with no uncertainty.
This trend continues until the number of background events
is comparable with the signal.
When the number of background events is comparable or larger than
the signal, the upper limit obtained is less stringent than
that extracted without including the Poisson uncertainty in the
estimated background.
It is therefore important to incorporate the Poisson uncertainty
into the upper limit; otherwise the upper limit obtained
could be too stringent.

\begin{table}[t]
\begin{center}
\caption[]{Upper limits on the signal for a few observed events
and various backgrounds in the one-dimensional case.
The total width of the sideband
regions is twice that of the signal region.}
\vspace{0.1in}
\label{tab:1d}
\begin{tabular}{ccccc}
\hline
& 
\multicolumn{2}{c}{90\%~C.L.} & 
\multicolumn{2}{c}{95\%~C.L.}\\
($N_{ob}$,$\frac{N_{sb}}{2})$ & $\lambda_0$ & $\lambda$ 
                              & $\lambda_0$ & $\lambda$ \\
\hline                                           
(0,$\frac{N_{sb}}{2}$) &2.30&2.30   &3.00&3.00 \\
(1,0)                  &3.89&3.61   &4.74&4.47 \\
(1,$\frac{1}{2}$)      &3.51&3.42   &4.36&4.27 \\
(1,1)                  &3.27&3.27   &4.11&4.11 \\
(1,$1\frac{1}{2}$)      &3.11&3.16   &3.94&3.99 \\
(1,2)                  &3.00&3.07   &3.82&3.90 \\
(2,0)                  &5.32&4.94   &6.30&5.92 \\
(2,1)                  &4.44&4.36   &5.41&5.33 \\
(2,2)                  &3.88&3.96   &4.82&4.91 \\
(2,3)                  &3.52&3.68   &4.44&4.61 \\
(2,4)                  &3.29&3.47   &4.18&4.39 \\
(3,0)                  &6.68&6.26   &7.75&7.34 \\
(3,1)                  &5.71&5.53   &6.78&6.61 \\
(3,2)                  &4.93&4.96   &5.98&6.03 \\
(3,3)                  &4.36&4.53   &5.40&5.58 \\
(3,6)                  &3.48&3.74   &4.42&4.73 \\
\hline                                           
\end{tabular}
\end{center}
\end{table}

\begin{table}[t]
\begin{center}
\caption[]{Upper limits on the signal for zero or one observed
event and various backgrounds in the two-dimensional case.
The total area of both the sideband and corner-band
regions is four times that of the signal region.}
\vspace{0.1in}
\label{tab:2d1}
\begin{tabular}{ccccc}
\hline
& 
\multicolumn{2}{c}{90\%~C.L.} & 
\multicolumn{2}{c}{95\%~C.L.}\\
($N_{ob}$,$\frac{N_{sb}}{2}$,$\frac{N_{cb}}{4}$)
  & $\lambda_0$ & $\lambda$  & $\lambda_0$ & $\lambda$ \\ 
\hline                            
(0,$\frac{N_{sb}}{2}$,$\frac{N_{cb}}{4}$) &2.30&2.30 &3.00&3.00 \\
(1,0,$\frac{N_{cb}}{4}$)                  &3.89&3.61 &4.74&4.47 \\  
(1,$\frac{1}{2}$,0)                       &3.51&3.48 &4.36&4.33 \\ 
(1,$\frac{1}{2}$,$\frac{1}{4}$)           &3.67&3.51 &4.53&4.36 \\
(1,$\frac{1}{2}$,$\frac{1}{2}$)           &3.89&3.53 &4.74&4.38 \\ 
(1,$\frac{1}{2}$,1)                       &3.89&3.55 &4.74&4.40 \\  
(1,$\frac{1}{2}$,2)                       &3.89&3.57 &4.74&4.43 \\ 
(1,1,0)                                   &3.27&3.35 &4.11&4.20 \\
(1,1,$\frac{1}{4}$)                       &3.38&3.40 &4.22&4.25 \\
(1,1,$\frac{1}{2}$)                       &3.51&3.44 &4.36&4.29 \\
(1,1,1)                                   &3.89&3.49 &4.74&4.33 \\
(1,1,2)                                   &3.89&3.53 &4.74&4.38 \\
(1,1,4)                                   &3.89&3.57 &4.74&4.42 \\
(1,2,0)                                   &3.00&3.14 &3.82&3.98 \\
(1,2,$\frac{1}{4}$)                       &3.05&3.21 &3.88&4.05 \\
(1,2,$\frac{1}{2}$)                       &3.11&3.26 &3.94&4.11 \\
(1,2,1)                                   &3.27&3.35 &4.11&4.19 \\
(1,2,2)                                   &3.89&3.44 &4.74&4.29 \\
(1,2,4)                                   &3.89&3.51 &4.74&4.36 \\
\hline                                           
\end{tabular}
\end{center}
\end{table}

\begin{table}[t]
\begin{center}
\caption[]{Upper limits on the signal for two observed
events and various backgrounds in the two-dimensional case.
The total area of both the sideband and corner-band
regions is four times that of the signal region.}
\vspace{0.1in}
\label{tab:2d2}
\begin{tabular}{ccccc}
\hline
& 
\multicolumn{2}{c}{90\%~C.L.} & 
\multicolumn{2}{c}{95\%~C.L.}\\
($N_{ob}$,$\frac{N_{sb}}{2}$,$\frac{N_{cb}}{4}$)
 & $\lambda_0$ & $\lambda$  & $\lambda_0$ & $\lambda$ \\ 
\hline                            
(2,0,$\frac{N_{cb}}{4}$) &5.32&4.94 &6.30&5.92 \\
(2,1,0)                  &4.44&4.49 &5.41&5.46 \\
(2,1,$\frac{1}{2}$)      &4.84&4.64 &5.82&5.61 \\
(2,1,1)                  &5.32&4.72 &6.30&5.69 \\
(2,1,2)                  &5.32&4.80 &6.30&5.77 \\
(2,1,4)                  &5.32&4.86 &6.30&5.83 \\
(2,2,0)                  &3.88&4.09 &4.82&5.05 \\ 
(2,2,$\frac{1}{2}$)      &4.13&4.31 &5.08&5.27 \\ 
(2,2,1)                  &4.44&4.45 &5.41&5.43 \\ 
(2,2,2)                  &5.32&4.62 &6.30&5.60 \\ 
(2,2,4)                  &5.32&4.76 &6.30&5.74 \\ 
(2,2,6)                  &5.32&4.82 &6.30&5.79 \\ 
(2,4,0)                  &3.29&3.55 &4.18&4.48 \\
(2,4,$\frac{1}{2}$)      &3.39&3.73 &4.30&4.67 \\
(2,4,1)                  &3.52&3.91 &4.44&4.86 \\
(2,4,2)                  &3.88&4.20 &4.82&5.17 \\
(2,4,4)                  &5.32&4.52 &6.30&5.49 \\
(2,4,6)                  &5.32&4.66 &6.30&5.63 \\
\hline                                           
\end{tabular}
\end{center}
\end{table}

For a large number of observed events, we can use the common
assumption that the signal $N_0$ in Eq.~(\ref{eq:signal}) is normally
distributed (Gaussian) with the variance given by
\begin{eqnarray}
\label{eq:sd}
\sigma^2 = N_{ob} + \frac{1}{4}N_{sb} + \frac{1}{16}N_{cb},
\end{eqnarray}
where we set $\alpha = \frac{1}{2}$ and
$\beta = \frac{1}{4}$.
The upper limit on the number of signal events 
$\lambda_G$ is obtained by integrating from $N_0$ to
$\lambda_G$ so that the integrated area is $\delta$ of the
area integrated to $+\infty$.
For an unphysical signal, $N_0 < 0$, the integration should be
renormalized by setting $N_0 = 0$ to obtain a more conservative
limit~\cite{PDG}.
Tables~\ref{tab:1d_big}-\ref{tab:2d_big} show a comparison
of $\lambda_G$ and $\lambda$ for the case of $N_{ob} = N_{bg}$.
Due to the longer tail of the Poisson distribution, $\lambda$
is larger than $\lambda_G$.
The significance of the difference diminishes with larger $N_{ob}$.
For example, in the one-dimensional case, it decreases from
$\sim 10\%$ for $N_{ob} = 10$ events to $\sim 5\%$ for
$N_{ob} = 50$ events.
For an experiment with small systematic error,
Eq.~(\ref{eq:1d_1}) or (\ref{eq:2d2_1}) should be used to compute
the upper limit instead of using the Gaussian approximation.

\begin{table}[t]
\begin{center}
\caption[]{Upper limits on the signal for large number of
observed events with large background in the one-dimensional case.
The total width of the sideband
regions is twice that of the signal region.}
\vspace{0.1in}
\label{tab:1d_big}
\begin{tabular}{ccccccc}
\hline
& 
\multicolumn{3}{c}{90\%~C.L.} & 
\multicolumn{3}{c}{95\%~C.L.}\\
($N_{ob}$,$\frac{N_{sb}}{2})$ & $\lambda_0$ & $\lambda$ & $\lambda_G$ 
                              & $\lambda_0$ & $\lambda$ & $\lambda_G$\\
\hline                                           
(10,10)                &6.63 &7.24 &6.35   &8.08 &8.78 &7.59  \\
(20,20)                &8.75 &9.82 &8.98   &10.60&11.83&10.74 \\
(35,35)                &11.11 &12.69&11.88 &13.41&15.24&14.20 \\
(50,50)                &13.00&15.00&14.20  &15.66&17.98&16.97 \\
\hline                                           
\end{tabular}
\end{center}
\end{table}

\begin{table}[t]
\begin{center}
\caption[]{Upper limits on the signal for large number of
observed events with large background in the two-dimensional case.
The total area of both the sideband and corner-band
regions is four times that of the signal region.}
\vspace{0.1in}
\label{tab:2d_big}
\begin{tabular}{ccccccc}
\hline
& 
\multicolumn{3}{c}{90\%~C.L.} & 
\multicolumn{3}{c}{95\%~C.L.}\\
($N_{ob}$,$\frac{N_{sb}}{2}$,$\frac{N_{cb}}{4}$) 
& $\lambda_0$ & $\lambda$ & $\lambda_G$  
& $\lambda_0$ & $\lambda$ & $\lambda_G$\\ 
\hline                            
(10,20,10)               &6.63&8.61  &7.78 &8.08 &10.31&9.30 \\
(15,25,10)               &7.78&9.86  &8.98 &9.45 &11.82&10.74 \\
(20,30,10)               &8.75&10.93&10.04 &10.60&13.10&12.00 \\
\hline                                           
\end{tabular}
\end{center}
\end{table}

For completeness, we also listed in
Tables~\ref{tab:1d_h}-\ref{tab:2d2_h} the 90\% and
95\% confidence level upper limits on the signal 
for $\alpha = \beta = 1$ with
and without including the Poisson uncertainty in
the estimated background for small number of observed events
and various backgrounds.

\begin{table}[t]
\begin{center}
\caption[]{Upper limits on the signal for a few observed events
and various backgrounds in the one-dimensional case.
The total width of the sideband
regions is the same as that of the signal region.}
\vspace{0.1in}
\label{tab:1d_h}
\begin{tabular}{ccccc}
\hline
& 
\multicolumn{2}{c}{90\%~C.L.} & 
\multicolumn{2}{c}{95\%~C.L.}\\
($N_{ob}$,$N_{sb}$) & $\lambda_0$ & $\lambda$ 
                              & $\lambda_0$ & $\lambda$ \\
\hline                                           
(0,$N_{sb}$)           &2.30&2.30   &3.00&3.00 \\
(1,0)                  &3.89&3.51   &4.74&4.36 \\
(1,1)                  &3.27&3.27   &4.11&4.11 \\
(1,2)                  &3.00&3.11   &3.82&3.94 \\
(1,3)                  &2.84&3.00   &3.64&3.82 \\
(1,4)                  &2.74&2.91   &3.53&3.72 \\
(1,6)                  &2.62&2.78   &3.39&3.58 \\
(2,0)                  &5.32&4.75   &6.30&5.72 \\
(2,1)                  &4.44&4.32   &5.41&5.29 \\
(2,2)                  &3.88&4.01   &4.82&4.97 \\
(2,3)                  &3.52&3.77   &4.44&4.72 \\
(2,4)                  &3.29&3.59   &4.18&4.52 \\
(2,6)                  &3.01&3.33   &3.86&4.23 \\
(3,0)                  &6.68&5.99   &7.75&7.08 \\
(3,1)                  &5.71&5.43   &6.78&6.52 \\
(3,2)                  &4.93&4.99   &5.98&6.06 \\
(3,3)                  &4.36&4.63   &5.40&5.69 \\
(3,4)                  &3.97&4.35   &4.97&5.39 \\
(3,6)                  &3.48&3.93   &4.42&4.94 \\
\hline                                           
\end{tabular}
\end{center}
\end{table}

\begin{table}[t]
\begin{center}
\caption[]{Upper limits on the signal for zero or one observed
event and various backgrounds in the two-dimensional case.
The total area of the sideband regions is twice that of the
signal region while the total area of the corner-band
regions is the same as that of the signal region.}
\vspace{0.1in}
\label{tab:2d1_h}
\begin{tabular}{ccccc}
\hline
& 
\multicolumn{2}{c}{90\%~C.L.} & 
\multicolumn{2}{c}{95\%~C.L.}\\
($N_{ob}$,$N_{sb}$,$N_{cb}$) & $\lambda_0$ & $\lambda$  
                             & $\lambda_0$ & $\lambda$ \\ 
\hline                            
(0,$N_{sb}$,$N_{cb}$)                     &2.30&2.30 &3.00&3.00 \\
(1,0,$N_{cb}$)                            &3.89&3.51 &4.74&4.36 \\  
(1,1,0)                                   &3.27&3.38 &4.11&4.22 \\
(1,1,1)                                   &3.89&3.42 &4.74&4.27 \\
(1,1,2)                                   &3.89&3.44 &4.74&4.29 \\
(1,1,3)                                   &3.89&3.45 &4.74&4.30 \\
(1,1,4)                                   &3.89&3.46 &4.74&4.31 \\
(1,1,6)                                   &3.89&3.47 &4.74&4.32 \\
(1,2,0)                                   &3.00&3.27 &3.82&4.11 \\
(1,2,1)                                   &3.27&3.34 &4.11&4.18 \\
(1,2,2)                                   &3.89&3.38 &4.74&4.22 \\
(1,2,3)                                   &3.89&3.40 &4.74&4.25 \\
(1,2,4)                                   &3.89&3.42 &4.74&4.27 \\
(1,2,6)                                   &3.89&3.44 &4.74&4.29 \\
(1,3,0)                                   &2.84&3.18 &3.64&4.02 \\
(1,3,1)                                   &2.99&3.27 &3.82&4.11 \\
(1,3,2)                                   &3.27&3.32 &4.11&4.17 \\
(1,3,3)                                   &3.89&3.36 &4.74&4.20 \\
(1,3,4)                                   &3.89&3.38 &4.74&4.22 \\
(1,3,6)                                   &3.89&3.41 &4.74&4.25 \\
\hline                                           
\end{tabular}
\end{center}
\end{table}

\begin{table}[t]
\begin{center}
\caption[]{Upper limits on the signal for two observed
events and various backgrounds in the two-dimensional case.
The total area of the sideband regions is twice that of the
signal region while the total area of the corner-band
regions is the same as that of the signal region.}
\vspace{0.1in}
\label{tab:2d2_h}
\begin{tabular}{ccccc}
\hline
& 
\multicolumn{2}{c}{90\%~C.L.} & 
\multicolumn{2}{c}{95\%~C.L.}\\
($N_{ob}$,$N_{sb}$,$N_{cb}$) & $\lambda_0$ & $\lambda$  
                            & $\lambda_0$ & $\lambda$ \\ 
\hline                            
(2,0,$N_{cb}$)           &5.32&4.75 &6.30&5.72 \\
(2,1,0)                  &4.44&4.50 &5.41&5.47 \\
(2,1,1)                  &5.32&4.57 &6.30&5.55 \\
(2,1,2)                  &5.32&4.61 &6.30&5.59 \\
(2,1,3)                  &5.32&4.64 &6.30&5.61 \\
(2,1,4)                  &5.32&4.65 &6.30&5.63 \\
(2,1,6)                  &5.32&4.68 &6.30&5.65 \\
(2,2,0)                  &3.88&4.29 &4.82&5.26 \\ 
(2,2,1)                  &4.44&4.41 &5.41&5.39 \\ 
(2,2,2)                  &5.32&4.49 &6.30&5.46 \\ 
(2,2,3)                  &5.32&4.53 &6.30&5.51 \\ 
(2,2,4)                  &5.32&4.56 &6.30&5.54 \\ 
(2,2,6)                  &5.32&4.61 &6.30&5.58 \\ 
(2,3,0)                  &3.52&4.10 &4.44&5.07 \\
(2,3,1)                  &3.88&4.27 &4.82&5.24 \\
(2,3,2)                  &4.44&4.37 &5.41&5.34 \\
(2,3,3)                  &5.32&4.43 &6.30&5.41 \\
(2,3,4)                  &5.32&4.48 &6.30&5.45 \\
(2,3,6)                  &5.32&4.54 &6.30&5.52 \\
(2,4,0)                  &3.29&3.94 &4.18&4.90 \\
(2,4,1)                  &3.52&4.14 &4.44&5.10 \\
(2,4,2)                  &3.88&4.26 &4.82&5.23 \\
(2,4,3)                  &4.44&4.34 &5.41&5.31 \\
(2,4,4)                  &5.32&4.40 &6.30&5.37 \\
(2,4,6)                  &5.32&4.48 &6.30&5.45 \\
\hline                                           
\end{tabular}
\end{center}
\end{table}

As noted in the previous section, Eq.~(\ref{eq:1d}) can also be used
to calculate the upper limit on the signal for the case in which the
background is estimated from a Monte Carlo simulation.
The impact of including the Poisson uncertainty in the estimated
background into the upper limit can be investigated by comparing the
limit with that obtained without including the uncertainty.
Table~\ref{tab:MC1} lists the 90\% confidence
level upper limits obtained with and without including the Poisson
uncertainty for several cases of small number of observed events
and various backgrounds.
As in the previous examples, $\lambda$ is smaller than $\lambda_0$
for small number of observed events with
comparable or smaller background.
However, $\lambda$ is larger than $\lambda_0$ for larger background.
Not including the Poisson uncertainty leads to a
too stringent limit in this case.

\begin{table}[t]
\begin{center}
\caption[]{90\% confidence level upper limits on the signal
for small number of observed events and various backgrounds
estimated from a Monte Carlo Simulation.}
\vspace{0.1in}
\label{tab:MC1}
\begin{tabular}{ccccccc}
\hline
$\alpha^{-1}$ 
				& \multicolumn{2}{c}{3} 
				& \multicolumn{2}{c}{6} 
    & \multicolumn{2}{c}{9} \\
\hline

($N_{ob}$,$N_{sb}$) & $\lambda_0$ & $\lambda$  
                    & $\lambda_0$ & $\lambda$ 
		    & $\lambda_0$ & $\lambda$ \\
\hline                         
(1,0) & 3.89&3.67 & 3.89&3.76 & 3.89&3.80 \\
(1,1) & 3.61&3.51 & 3.74&3.65 & 3.79&3.71 \\
(1,2) & 3.42&3.38 & 3.61&3.55 & 3.70&3.64 \\
(1,3) & 3.27&3.27 & 3.51&3.47 & 3.61&3.57 \\
(1,4) & 3.16&3.18 & 3.42&3.40 & 3.54&3.51 \\
(1,5) & 3.07&3.11 & 3.34&3.33 & 3.48&3.45 \\
(1,6) & 3.00&3.05 & 3.27&3.27 & 3.42&3.40 \\
(1,7) & 2.93&3.00 & 3.21&3.22 & 3.37&3.36 \\
(1,8) & 2.88&2.95 & 3.16&3.17 & 3.32&3.31 \\
(1,9) & 2.84&2.91 & 3.11&3.13 & 3.27&3.27 \\
(2,0) & 5.32&5.04 & 5.32&5.17 & 5.32&5.22 \\
(2,1) & 5.00&4.79 & 5.16&5.02 & 5.21&5.11 \\
(2,2) & 4.70&4.57 & 5.00&4.88 & 5.10&5.01 \\
(2,3) & 4.44&4.38 & 4.84&4.75 & 5.00&4.92 \\
(2,4) & 4.22&4.21 & 4.70&4.63 & 4.89&4.82 \\
(2,5) & 4.04&4.07 & 4.57&4.51 & 4.79&4.73 \\
(2,6) & 3.88&3.94 & 4.44&4.41 & 4.70&4.65 \\
(2,7) & 3.74&3.82 & 4.33&4.31 & 4.61&4.57 \\
(2,8) & 3.62&3.72 & 4.22&4.22 & 4.52&4.49 \\
(2,9) & 3.52&3.63 & 4.13&4.13 & 4.44&4.42 \\
\hline                                           
\end{tabular}
\end{center}
\end{table}

\section{Conclusion}

We have presented a procedure for calculating an upper limit on
the number of signal events
which incorporates the Poisson uncertainty in the background
estimated from control regions of one or two dimensions. 
For small number of observed events in the signal region,
the limit obtained is more stringent than that extracted
assuming no uncertainty in the estimated background.
This trend continues until the number of background events
is comparable with the signal.
When the number of background events is comparable or larger than
the signal, the upper limit obtained is less stringent than that
extracted without including the Poisson uncertainty in the
estimated background.
It is therefore important to incorporate the Poisson uncertainty
into the upper limit; otherwise the upper limit obtained
could be too stringent.

\section{Acknowledgment}
This work was supported in part by the U.S.~Department of Energy.
K.K.G. thanks the OJI program of DOE for support.



\end{document}